\documentclass{PoS}

\PoS{PoS(BDMH2004)}

\usepackage{epsfig}

\title{Using globular clusters to test gravity in the weak acceleration regime: NGC 6171}

\ShortTitle{Using globular clusters to test gravity in the weak acceleration regime: NGC 6171}

\author{\speaker{Riccardo Scarpa}, Gianni Marconi, and Roberto Gilmozzi\\
        European Southern Observatory, Chile\\
        E-mail: \email{rscarpa@eso.org, gmarconi@eso.org, rgilmozz@eso.org }}

\abstract{ As part of an ongoing program to test Newton's law of
gravity in the low acceleration regime using globular clusters, we
present here new results obtained for NGC 6171.  Combining VLT spectra
for 107 stars with data from the literature, we were able to trace the
velocity dispersion profile up to $\sim$ 16 pc from the cluster
center, probing accelerations of gravity down to $3.5 \times 10^{-9}$ cm s$^{-2}$. 
The velocity dispersion is found to remain constant at large
radii (with an asymptotic values of $2.7$ km s$^{-1}$) rather than
follow the Keplerian falloff. Similar results were previously found
for the globular clusters $\omega$ Centauri and M15.  We have now
studied three clusters and all three have been found to have a flat
dispersion profile beyond the radius where their internal acceleration
of gravity is $\sim a_0=1.2 \times 10^{-8}$ cm s$^{-2}$.  Whether this
indicates a failure of Newtonian dynamics or some more conventional
dynamical effect (e.g., tidal heating) is still unclear. However, the
similarities emerging between globular clusters and elliptical
galaxies seem to favor the first of the two possibilities.}

\FullConference{Baryons in Dark Matter Halos\\
		 5-9 October 2004\\
		 Novigrad, Croatia}

\begin{document}
\section{Introduction}
We present here results of an ongoing experiment designed to test the
validity of Newton's law of gravity in the low acceleration regime.
The idea sparked from the consideration that the typical acceleration
governing the dynamics of stellar structures is much smaller than the
smallest acceleration probed in our laboratories or in the solar
system.  Thus, any time Newton's law is applied to galaxies, for
instance to infer the existence of non-baryonic dark matter (hereafter
DM), its validity is extrapolated by several orders of
magnitude. Interestingly, unanimous agreement has been reached (e.g.,
\cite{binney04}) on the fact that deviations from Newtonian dynamic
are $always$ observed when and only when the
gravitational acceleration falls below a certain value ($\sim 10^{-8}$
cm s$^{-2}$, as computed considering only baryons).  This systematic
behavior suggests we may be facing a breakdown of Newton's law rather than the
effects of a still undetected medium. This fact is at the foundation
of a particular modification of Newtonian dynamics known as MOND
(\cite{milgrom83}), which postulates a breakdown of Newton's law of
gravity (or inertia) below $a_0=1.2\times 10^{-8}$ cm s$^{-2}$
(\cite{begeman91}). MOND successfully describes the properties of an
increasingly large number of stellar structures without invoking DM
(\cite{mcgaugh98}; \cite{sanders02}).  Unfortunately, in the realm of
galaxies MOND and DM provides alternative though indistinguishable
descriptions of the data, making it difficult to decide in favor of one
of the two options.

Irrespectively of the validity of MOND, to discriminate between a
failure of Newton's law and DM one as to perform an experiment known
{\it a priori} to be free from the effects of DM. For instance, if
deviations from Newtonian dynamic were to be observed in the
laboratory these will have to be ascribed to a breakdown of Newtonian
dynamics rather than to the effects of DM.  Interestingly, there is
general agreement that, if any, the effects of DM on globular clusters
are dynamically negligible, making them perfect for testing Newtonian
dynamics down to arbitrarily small accelerations.  In two previous
papers \cite{scarpa03A} \cite{scarpa03B} we presented a study of the
dynamical properties of the globular clusters $\omega$ Centauri and
M15, in which accelerations below $a_0$ are probed. It was shown that
as soon as the acceleration reaches this value the velocity dispersion
remains constant instead of following the Keplerian falloff
(Fig. \ref{omegacen}).  This result is similar to what is
observed in galaxies and explained invoking DM, something we can not
do in the present case. In an attempt to generalize this result, we
report here new results for NGC 6171, a compact globular cluster
located 6.4 kpc from the sun and 3.3 kpc from the galactic center
(\cite{harris96}).

\begin{table}[b]
\caption{Radial velocity dispersion for NGC 6171}
\begin{tabular}{c|ccc|ccc}
Bin limits  & Stars/bin & bin center & $\sigma$ (km s$^{-1}$)& Stars/bin & 
bin center & $\sigma$ (km s$^{-1}$)\\
\hline
~&\multicolumn{3}{c}{VLT data} & \multicolumn{3}{c}{VLT plus Piatek data}\\
\hline
$0-2    $&    &      &                  & 28 &  0.9  & 4.03 $\pm$ 0.36\\
$2-5    $& 30 &  3.8 & 3.25  $ \pm$ 0.28& 54 &  3.7  & 3.24 $\pm$ 0.18\\
$5-8    $& 35 &  6.5 & 2.53  $ \pm$ 0.19& 46 &  6.5  & 2.82 $\pm$ 0.18\\
$8-13   $& 23 & 10.4 & 2.40  $ \pm$ 0.25& 27 & 10.3  & 2.69 $\pm$ 0.25\\
$13-23  $& 20 & 16.7 & 2.68  $ \pm$ 0.31& 20 & 16.7  & 2.68 $\pm$ 0.31\\
\hline
\hline
\multicolumn{3}{l}{Distances in parsecs.}
\end{tabular}
\end{table}

\section{Observation and results for NGC 6171}

Data for NGC 6171 were obtained at the ESO VLT telescope using FLAMES,
a fiber multi-objects spectrograph. Two sets of stars, selected based
on color, were prepared. One containing red-giant branch stars, an the
other fainter stars down to the turn off of the main sequence.  In
total spectra for $\sim 170$ were obtained in the wavelength range
5143-5356 \AA\ at resolution $\sim$ 25900.  Data reduction was
performed with the FLAMES pipeline and radial velocities derived using
Fourier cross correlation techniques.  In most cases, an accuracy
better that 0.75 km s$^{-1}$ was achieved in the radial velocity. Due
to the small dispersion we are trying to measure ($\sim 2-3$ km
s$^{-1}$), we decided to consider only those stars for which a final
accuracy better that 1.25 km s$^{-1}$ was achieved. This left us with
118 stars.

\begin{figure}
\centering
\includegraphics[height=7cm]{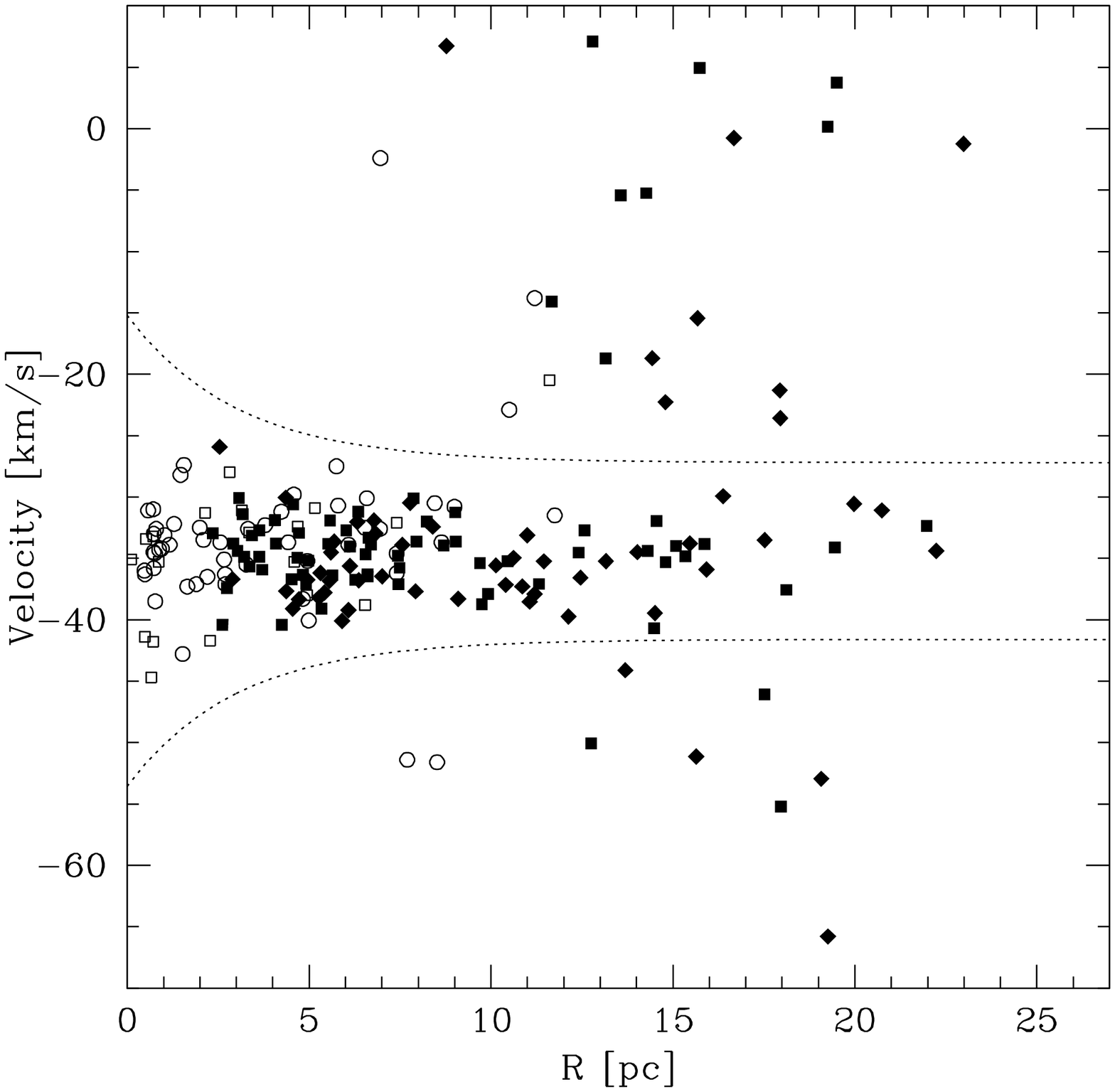}
\includegraphics[height=7cm]{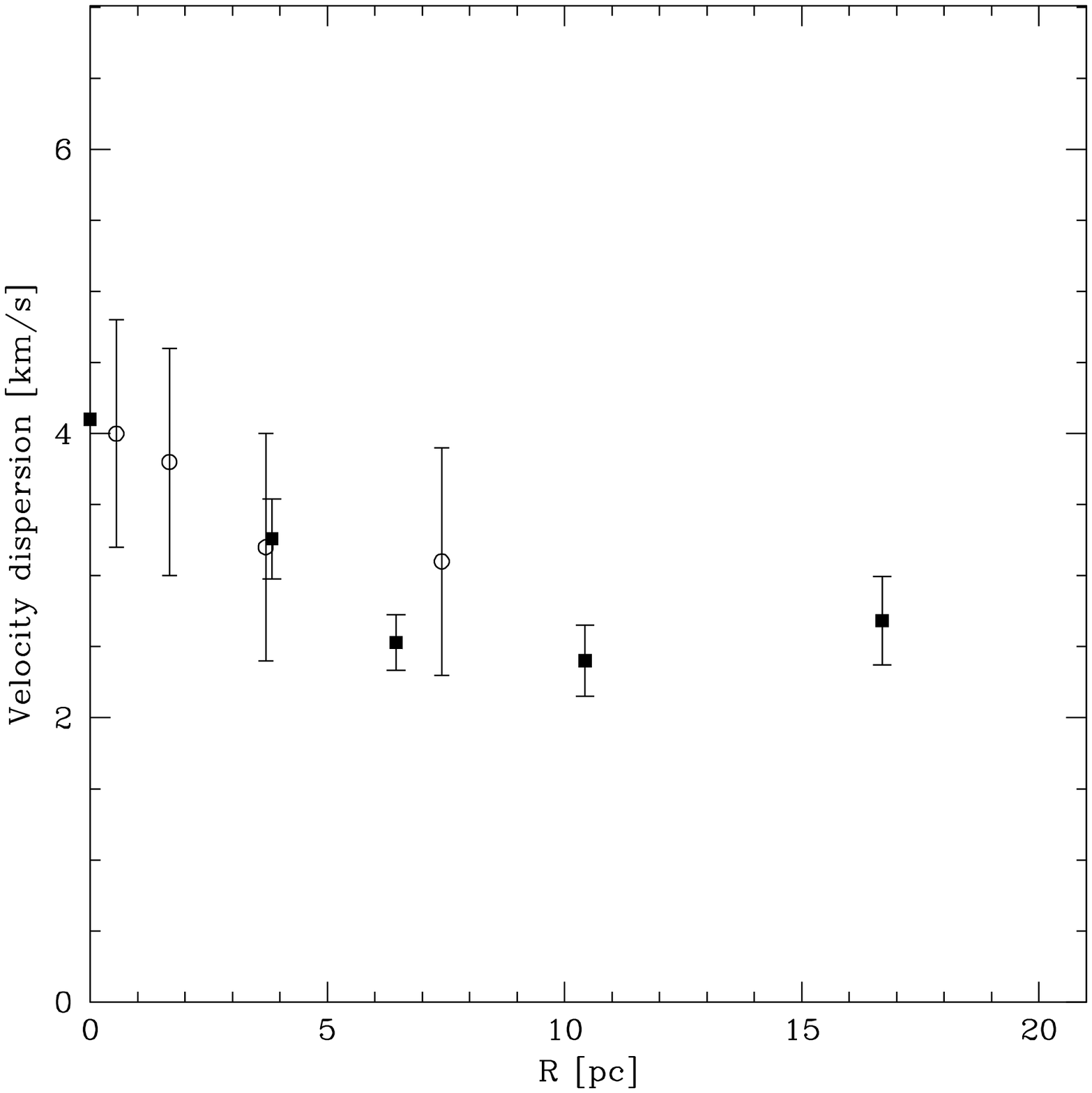}
\caption{\label{ngc6171_vel} {\bf Left:} Radial velocities as a
function of distance from the cluster center.
Squares and diamonds refer to stars from the bright and faint dataset,
respectively.  Only data with accuracy better than 1.25 km s$^{-1}$
are shown.  Data from \cite{piatek94} are plotted as  open symbols. Circles
and squares refer to values with uncertainty smaller and greater than 1.25
km s$^{-1}$, respectively. The two dotted lines define the
region used to select cluster members. {\bf Right:} Comparison of the
velocity dispersion as derived using our data (Squares) and data
published by \cite{piatek94} (Circles).  Within
uncertainties (error-bars represent 1$\sigma$ uncertainties) there is
good agreement between the two datasets. The central velocity
dispersion of 4.1 km s$^{-1}$ is from \cite{harris96}.  }
\end{figure}

A crucial aspect of this study is the membership determination.  This
is important in the present case because of the low
galactic latitude of NGC 6171 and its radial velocity of only -33.6 km
s$^{-1}$ (\cite{harris96}).  In Fig. \ref{ngc6171_vel} the locus of
cluster members is immediately identified as an over-density in the
parameter space, in particular for stars closer than 10 pc from the
center.  To identify the members outward of 10 pc, we used data in the
region between 6 and 9 pc from the center, which have velocity
dispersion $\sigma=2.4$ km s$^{-1}$.  We then define as members all
stars with velocity falling within $\pm 3\sigma$ from the average.
This procedure left us with 107 stars that, notably, are all well
within the cluster tidal radius (32.5 pc, \cite{harris96}). The
corresponding velocity dispersion is given in Tab. 1 and 
compared to data from the literature in Fig. \ref{ngc6171_vel}.

\section{Discussion}

\begin{figure}
\centering
\includegraphics[height=5cm]{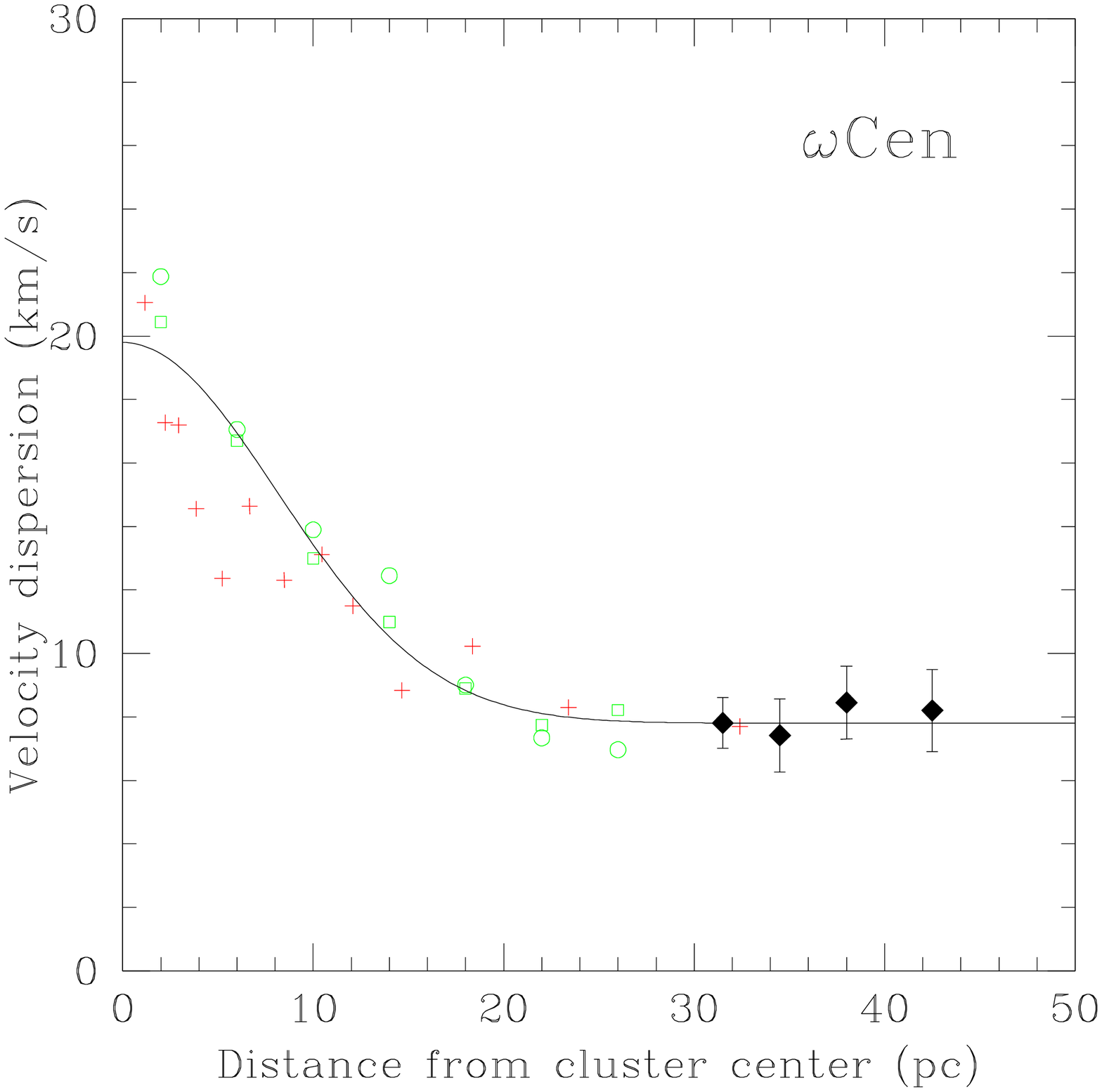}\includegraphics[height=5cm]{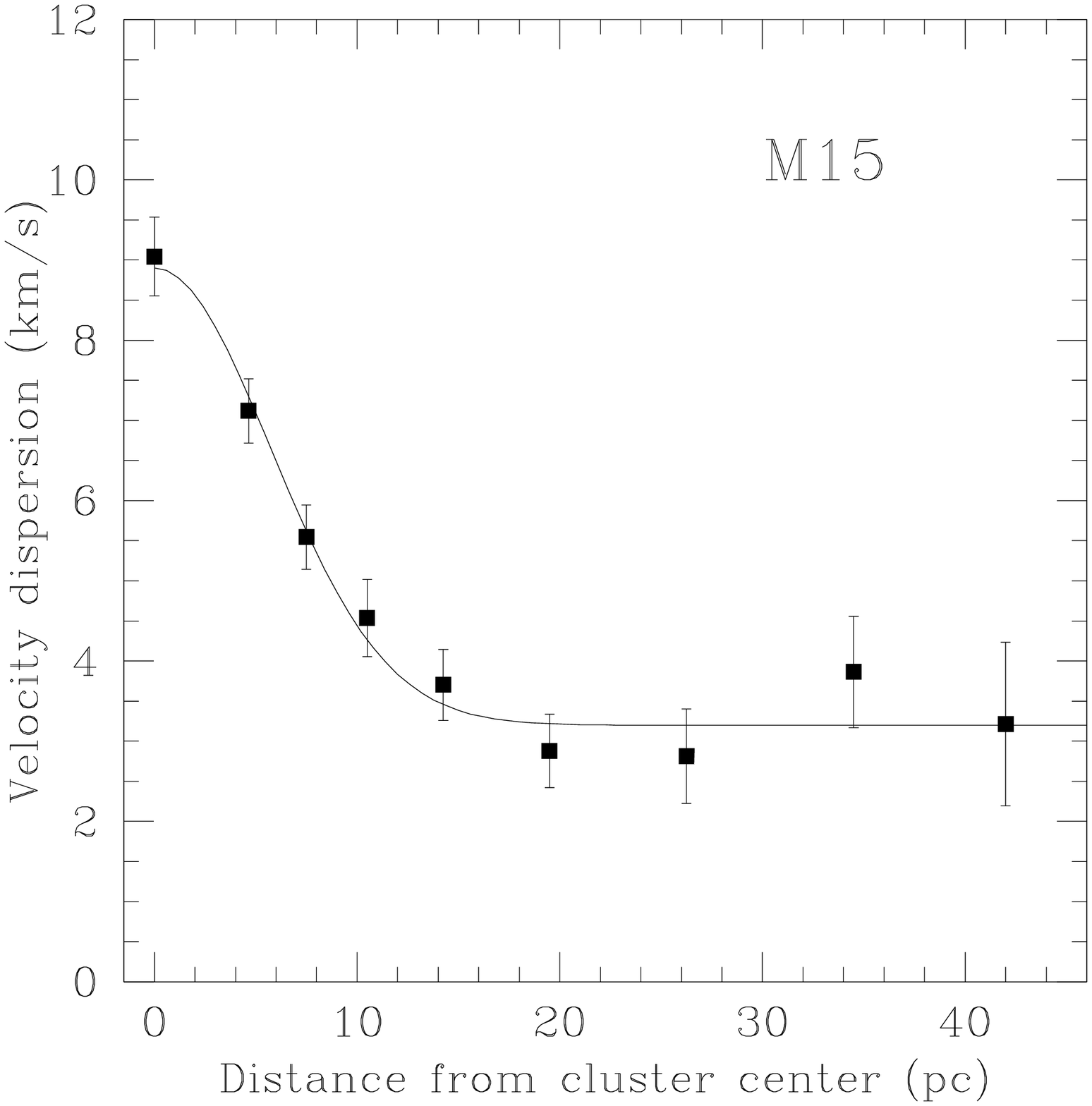}
\includegraphics[height=5cm]{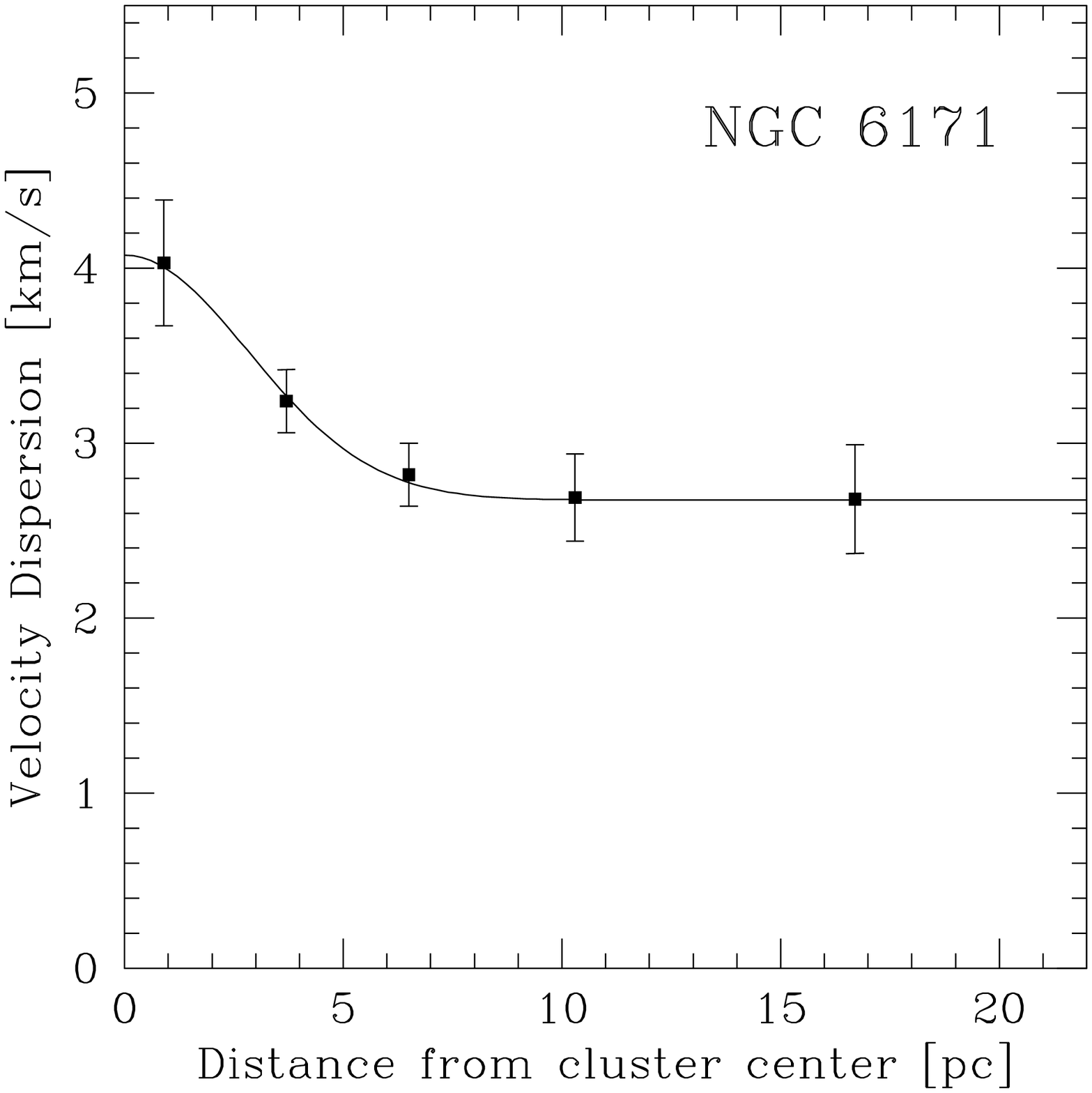}
\caption{\label{omegacen} {\bf Left:} The velocity dispersion
profile of $\omega$ Centauri (as presented in \cite{scarpa03A})
flattens out at R$=27\pm 3$ pc, where $a = 2.1\pm 0.5 \times 10^{-8}$ cm s$^{-2}$.
{\bf Center:} Dispersion profile for M15 as derived from data
by \cite{drukier98}. The flattening occurs at R$=18\pm 3$ pc,
equivalent to $a = 1.7\pm 0.6 \times 10^{-8}$ cm s$^{-2}$. 
{\bf Right:}  Velocity dispersion profile for NGC
6171 as derived combining our VLT data with data from \cite{piatek94}.
The profile remains flat within uncertainties outward of $8\pm1.5$ pc, where 
 $a=1.4^{+0.7}_{-0.4}\times 10^{-8}$ cm s$^{-2}$.
In all panels, the solid line is a fit obtained using a Gaussian plus a constant,
ment to better evintiate the flattening of the profile.}
\end{figure}

We have presented new data for NGC 6171, bringing to three the number
of globular clusters for which accelerations below $a_0$ have been probed.
In spite of having different physical properties as well as different
dynamical and evolutionary histories, all three clusters behave
exactly in the same way, both quantitatively and qualitatively
(Fig. \ref{omegacen}).  Assuming a mass-to-light ratio of one in solar
units, the flattening of the dispersion profile occurs for very
similar values of the internal acceleration of gravity, with an
average value of $a = 1.78\pm 0.4 \times 10^{-8}$, very close to the
MOND prediction $a_0$.

While a striking similarity with elliptical galaxies is emerging, it
is still possible to find explanations for this behavior within the
boundaries of Newtonian dynamics (tidal heating would be a possible
one), though more and more fine tuning is necessary to explain all
these ``coincidences''.

\end{document}